\documentstyle[seceq,epsf]{ptptex}
\notypesetlogo  %comment in if to eliminate PTPTeX logo
\title{%        %You can use \\ for explicit line-break
Generation of Post-Newtonian Gravitational Radiation \\
via Direct Integration of the Relaxed Einstein Equations
}
\author{%       %Use \sc for the family name
Clifford M. {\sc Will}\footnote{E-mail address: cmw@wuphys.wustl.edu}
}

\inst{%         %Affiliation, neglected when [addenda] or [errata]
Department of Physics and McDonnell Center for the Space Sciences
\\
Washington University, St. Louis MO 63130 U.S.A }

\recdate{%      %Editorial Office will fill in this.
\today
}

\abst{%       %this abstract is neglected when [addenda] or [errata]
The completion of a network of advanced laser-interferometric
gravitational-wave observatories 
around 2001
will make possible the study of the inspiral and coalescence of binary
systems of compact objects (neutron stars and black holes), using
gravitational radiation. To extract useful information from the waves,
such as the masses and spins of the bodies, theoretical general
relativistic gravitational waveform templates of extremely high
accuracy will be needed for filtering the data,
probably as accurate as $O[(v/c)^6]$ beyond the predictions of the
quadrupole formula.  
We summarize a method, called DIRE, for Direct Integration of the
Relaxed Einstein Equations, which
extends and improves an earlier framework due to Epstein and Wagoner,
in which Einstein's equations are recast as a flat
spacetime wave equation with source composed of matter confined to
compact regions and gravitational non-linearities extending to
infinity.  The new method
is free of divergences or undefined
integrals, correctly predicts all gravitational wave ``tail'' effects
caused by backscatter of the outgoing radiation off the background
curved spacetime, and yields radiation that propagates asymptotically
along true null cones of the curved spacetime.  The method also yields
equations of motion through $O[(v/c)^4]$, radiation-reaction terms at
$O[(v/c)^5]$ and $O[(v/c)^7]$, and gravitational waveforms and energy
flux through $O[(v/c)^4]$, in agreement with other approaches.  We
report on progress in evaluating the $O[(v/c)^6]$ contributions.
}

\begin{document}

\maketitle

\section{Introduction}

Some time in the next decade, a new window for astronomy and
relativistic gravity may be realized, with
the completion and operation of kilometer-scale, laser interferometric
gravitational-wave observatories in the U.S. (LIGO project), Europe
(VIRGO and GEO-600 projects) and Japan (TAMA 300 project).  
Gravitational-wave searches at these observatories are scheduled to
commence around 2002.  The LIGO 
broad-band antennae will have the capability of detecting and
measuring the gravitational waveforms from astronomical sources in a
frequency band between about 10 Hz (the seismic noise cutoff) and 
500 Hz (the photon counting noise cutoff), with a maximum
sensitivity to strain at around 100 Hz of $\Delta l/l \sim 10^{-22}$
(rms).  The most promising source for detection and study of
the gravitational-wave signal is the ``inspiralling compact binary''
-- a binary system of neutron stars or black holes (or one of each) in
the final minutes of a death dance leading to a violent merger.
Such is the fate, for example of the Hulse-Taylor binary pulsar PSR
1913+16 in about 240 million years.  Given the expected sensitivity of the
``advanced LIGO'' (around 2007), which could see such sources out to
hundreds of megaparsecs, it has been estimated that from 3 to
100 annual inspiral events could be detectable.
Other sources, such as supernova core collapse events, instabilities
in rapidly rotating nascent neutron stars, signals from
non-axisymmetric pulsars, and a stochastic background of waves, may be
detectable (for reviews, see Ref. \citen{snowmass} and other articles in this
volume).  

The analysis of gravitational-wave data from such inspiral sources will involve
some form of matched filtering of the noisy detector output against an
ensemble of theoretical ``template'' waveforms which depend on the
intrinsic parameters of the inspiralling binary, such as the component
masses, spins, and so on, and on its inspiral evolution.  
How accurate must a template be in order to
``match'' the waveform from a given source (where by a match we mean
maximizing the cross-correlation or the 
signal-to-noise ratio)?  In the total accumulated phase
of the wave detected in the sensitive bandwidth, the template must
match the signal to a fraction of a cycle.  For two inspiralling
neutron stars, around 16,000 cycles should be detected; this implies a
phasing accuracy of $10^{-5}$ or better.  Since $v/c \sim 1/10$ during
the late inspiral, this means that correction terms in the phasing
at the level of
$(v/c)^5$ or higher are needed.  More formal analyses confirm this
intuition.\cite{finnchern}  

Because it is a slow-motion system ($v/c \sim 10^{-3}$), the binary
pulsar is sensitive only to the lowest-order effects of gravitational
radiation as predicted by the quadrupole formula.  Nevertheless, the
first correction terms of order $v/c$ and $(v/c)^2$ to the quadrupole formula,
were
calculated as early as 1976 \cite{wagwill}.  These are now
conventionally called ``post-Newtonian'' (PN) corrections, with each power
of $v/c$ corresponding to half a post-Newtonian order (0.5PN), in analogy with
post-Newtonian corrections to the Newtonian
equations of motion.\cite{convention} \ In 1976, the post-Newtonian
corrections to the quadrupole formula were of purely academic, rather
than observational interest.  

But for laser-interferometric observations of gravitational waves,
the bottom line is that, in order to measure the astrophysical
parameters of the source and to test the properties of the
gravitational waves, it is necessary to derive the
gravitational waveform and the resulting radiation back-reaction on the orbit
phasing at least to 2PN,
or second post-Newtonian order, $O[(v/c)^4]$, beyond the quadrupole
approximation, and probably to 3PN
order.  

\section{Post-Newtonian Generation of Gravitational Waves}

The motion of isolated binary systems and the
generation of gravitational radiation are long-standing problems
that date back to the first years following the publication of
GR, when Einstein calculated the gravitational
radiation emitted by a laboratory-scale object using the linearized
version of GR.
Shortly after the discovery of the binary pulsar PSR
1913+16 in 1974, questions were raised about the foundations of the
``quadrupole formula'' for gravitational radiation damping 
(and in
some quarters, even about its quantitative validity).
These questions were answered in part by theoretical 
work designed to shore up the
foundations of the quadrupole approximation,\cite{walkerwill} and in part 
(perhaps mostly) by the
agreement between the predictions of the
quadrupole formula and the {\it observed} 
rate of damping of the pulsar's orbit.

The challenge of providing accurate templates for LIGO-VIRGO data
analysis has led to major efforts to calculate gravitational waves to
high PN order.  Three approaches have been developed.

The BDI approach of Blanchet, Damour and Iyer is 
based on a mixed post-Newtonian and 
``post-Minkowskian'' framework for
solving Einstein's equations approximately, developed in a series of
papers by Damour and colleagues.\cite{bd86}  
The idea is to solve
the vacuum Einstein equations in the exterior of the
material sources extending out to the radiation zone
in an expansion (``post-Minkowskian'')
in ``nonlinearity'' (effectively an
expansion in powers of Newton's constant $G$), and to express the
asymptotic solutions in terms of a set of formal, time-dependent, 
symmetric and trace-free (STF) multipole moments.\cite{thorne80}  
Then, in a near
zone within one characteristic wavelength of the radiation, the
equations including the material 
source are solved in a slow-motion approximation (expansion in powers
of $1/c$)
that yields a set of STF source multipole moments expressed as integrals over 
the ``effective'' source, including both matter and 
gravitational field contributions.  
The solutions involving the
two sets of moments are then matched in an intermediate zone, resulting
in a connection between the formal radiative moments and the source moments.
The matching also provides a natural way, using analytic continuation,
to regularize integrals involving the non-compact contributions of
gravitational stress-energy, that might otherwise be divergent.  For
further details, see the article by Blanchet in this volume. 

An approach called DIRE is based on a framework developed by
Epstein and Wagoner (EW)\cite{ew}, and extended by Will, Wiseman and Pati.  
We shall describe DIRE briefly below.

A third approach, valid only 
in the limit in which one mass is much smaller than the other,
is that of black-hole perturbation theory.  This method provides 
numerical results that are exact in $v/c$, as well as analytical results 
expressed as series in powers of $v/c$, both for
non-rotating and for rotating black holes.  For
non-rotating holes, the analytical expansions have been carried to
{\it 5.5} PN order.\cite{tts96}  In all cases of suitable overlap, the
results of all three methods agree precisely.

\section{Direct Integration of the Relaxed Einstein Equations
(DIRE)}

Like the BDI approach,
DIRE involves rewriting the Einstein
equations in their ``relaxed'' form, namely as an inhomogeneous,
flat-spacetime wave equation for a field $h^{\alpha\beta}$, 
whose formal solution can be written
\begin{eqnarray}
h^{\alpha \beta} (t,{\bf x}) = && 4 \int_{\cal C}
{ \tau^{\alpha \beta} (t -| {\bf x} - {\bf x^\prime} |, {\bf x^\prime}
)
\over | {\bf x} - {\bf x^\prime} | } d^3x^\prime \;,
\label{nearintegral}
\end{eqnarray}
where the 
source $\tau^{\alpha\beta}$
consists of both the material stress-energy, and a
``gravitational stress-energy'' made up of all the terms non-linear in
$h^{\alpha\beta}$, and the integration is over the past flat-spacetime
null cone $\cal C$ of the field point $(t,{\bf x})$ (see Fig. 1).  The 
wave equation is accompanied by a harmonic or
deDonder gauge condition
$h^{\alpha\beta},_{\beta} = 0$, which serves to specify a coordinate system, and
also imposes equations of motion on the sources.  Unlike the BDI
approach, a {\it single} formal solution is written down, valid everywhere
in spacetime.  This formal solution,
is then iterated in a
slow-motion ($v/c<1$), weak-field ($||h^{\alpha\beta}|| <1$ )
approximation, that is very similar to the corresponding procedure in
electromagnetism.  However, because the integrand of this retarded integral   
is not compact by virtue of the non-linear field contributions, the
original EW formalism quickly runs up against integrals that are not
well defined, or worse, are divergent.  Although at the lowest
quadrupole and first few PN orders, various arguments can be given to
justify sweeping such problems under the rug,\cite{wagwill} they are not very
rigorous, and provide no guarantee that the divergences do not become
insurmountable at higher orders.  As a consequence, despite 
efforts to cure the problem, the EW formalism fell into
some disfavor as a route to higher orders, although an extension to
1.5PN order was accomplished.\cite{magnum}

\begin{figure}
\begin{center}
\leavevmode
\epsfysize=3.5in \epsffile{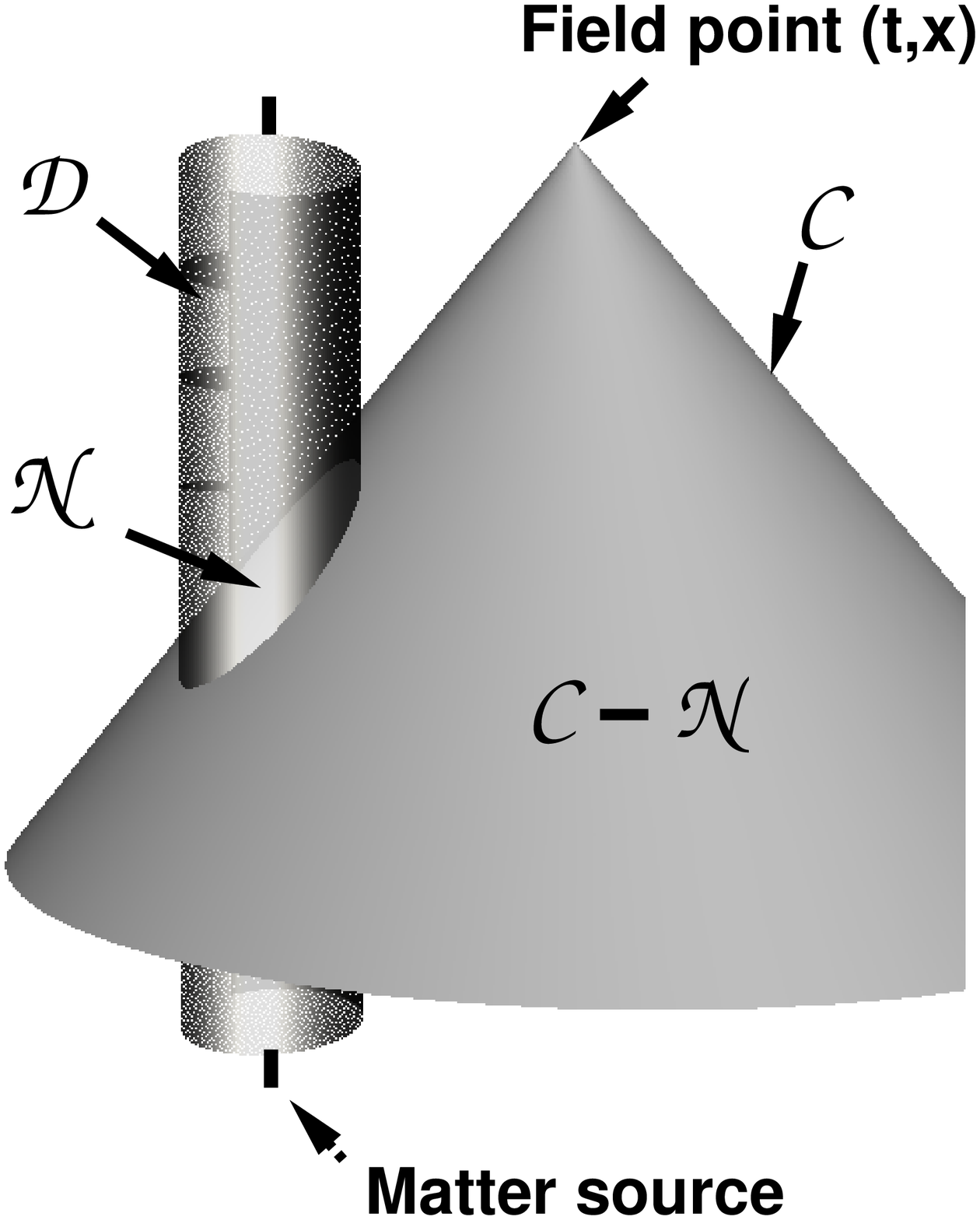}
\end{center}
\caption{Past harmonic null cone $\cal C$ of the field point $(t,{\bf
x})$ intersects the near zone $\cal D$ in the hypersurface $\cal N$.}
\begin{center}
\leavevmode
\epsfysize=3.5in \epsffile{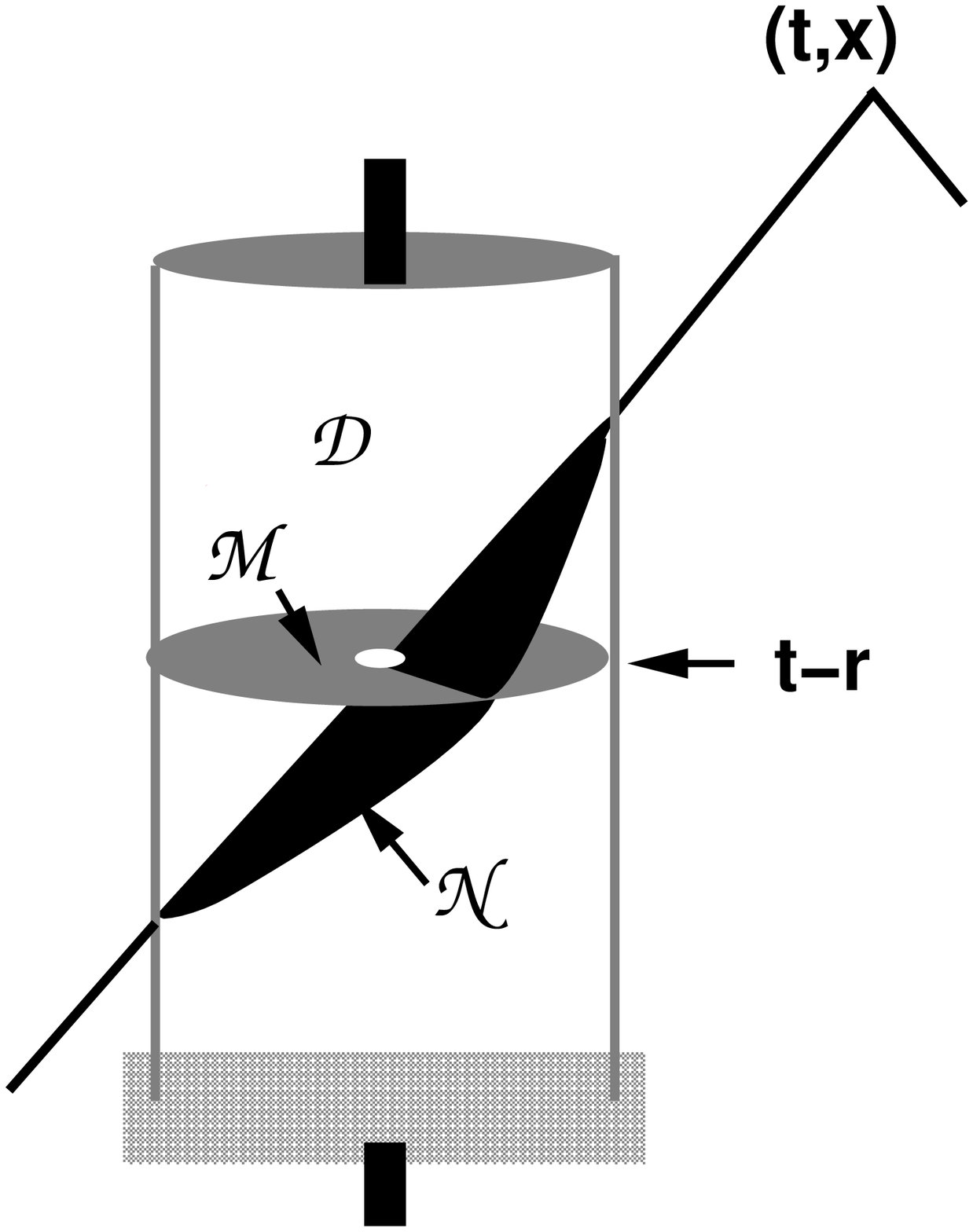}
\end{center}
\caption{Taylor expansion of retarded time dependence on $\cal N$
results in multipole moments integrated over the spatial hypersurface
$\cal M$, corresponding to fixed retarded time $t-r$}
\end{figure}

The resolution of this problem involves taking
literally the statement that the solution is a {\it retarded} integral,
{\it i.e.} an integral over the {\it entire} past null cone of the field
point.\cite{opus}  To be sure, that part of the integral that extends over
the intersection $\cal N$ between the past null cone and 
the material source and the near zone 
is still approximated as usual by a slow-motion expansion
involving spatial integrals over a constant-time hypersurface $\cal M$ of 
moments of the source, including the
non-compact gravitational contributions, just as in the BDI framework
(Fig. 2).  
But instead of cavalierly extending the
spatial integrals to infinity as was implicit in the original EW
framework, and risking undefined or 
divergent integrals, we terminate the integrals at
the boundary of the near zone, chosen to be at a radius $\cal R$ given
roughly by one wavelength of the gravitational radiation.  
{}For the 
integral over the rest of the past null cone
exterior to the near zone (``radiation zone''), we neither make a slow-motion
expansion nor continue to integrate over a spatial hypersurface, 
instead we use a coordinate transformation
in the integral
from the spatial coordinates $d^3x^\prime$ to quasi-null coordinates
$du^\prime$, $d\theta^\prime$, $d\phi^\prime$, where
\begin{equation}
t-u^\prime =r^\prime + |{\bf x} -{\bf x}^\prime | \,,
\end{equation}
to convert
the integral into a convenient, easy-to-calculate form, that is
manifestly convergent, subject only to reasonable assumptions about
the past behavior of the source:  
\begin{equation}
h_{{\cal C}-{\cal N}}^{\alpha\beta}(t,x)=
4 \int_{-\infty}^u du^\prime \int {{\tau^{\alpha\beta}
(u^\prime+r^\prime, {\bf x}^\prime )} \over {t-u^\prime-{\bf n}^\prime
\cdot {\bf x}} } [r^{\prime} (u^\prime, \Omega^\prime )]^2 d^2
\Omega^\prime \,.
\label{outer}
\end{equation}
This transformation was
suggested by earlier work on a non-linear gravitational-wave
phenomenon called the Christodoulou memory.\cite{christo} \ 
Not only are all integrations now
explicitly finite and convergent, one can show that all
contributions from the finite, near-zone spatial integrals that depend
upon
$\cal R$ are actually {\it cancelled} by corresponding terms from the
radiation-zone integrals, valid for both positive and negative powers
of $\cal R$ and for terms logarithmic in $\cal R$.\cite{willpati} \ 
Thus the procedure, as expected, has no
dependence on the artificially chosen boundary radius $\cal R$
of the near-zone.  In
addition, the method can be carried to higher orders in a
straightforward manner.  The result is a 
manifestly finite,
well-defined procedure for calculating gravitational radiation to high
orders.  

Thus, for field points in the far zone, the integral over the near
zone takes the standard form of a multipole expansion,
\begin{subeqnarray}
h_{\cal N}^{\alpha\beta}(t, {\bf x}) = 4 \sum_{m=0}^\infty {(-1)^m
\over
{m!}} {\partial^m \over {\partial x^{k_1} \dots \partial x^{k_m}}}
\left ( {1 \over r} M^{\alpha\beta k_1 \dots k_m}(t-r) \right ) \,,
\\
M^{\alpha\beta k_1  \dots k_m}(t-r) \equiv \int_{\cal M}
\tau^{\alpha\beta} (t-r,x^\prime)x^{k_1^\prime} \dots x^{k_m^\prime} 
d^3x^\prime \,,
\end{subeqnarray}
where integrals are over the {\it finite} hypersurface $\cal M$.  For
field points in the near zone, the integral over the near
zone can be expanded in the form of a sequence of ``Poisson''-like
potentials and superpotentials (and ``superduper''-potentials) 
evaluated at a fixed time $t$:
\begin{equation}
h_{\cal N}^{\alpha\beta}(t, {\bf x}) = 4 \sum_{m=0}^\infty {(-1)^m
\over
{m!}} {\partial^m \over {\partial t^m}}
\int_{\cal M}
\tau^{\alpha\beta} (t,x^\prime) |{\bf x}-{\bf x}^\prime |^{m-1}
d^3x^\prime  \,.
\end{equation}
In each case, the integrals must be combined with the corresponding
integral over the
rest of the past light cone, Eq. (\ref{outer}).  
Because of the aforementioned general
proof, it is not necessary to keep any terms in these integrals that
depend explicitly on the radius $\cal R$; this simplifies calculations
considerably.

\section{Equations of Motion to 3.5PN Order}

We assume that the orbiting bodies are sufficiently small compared to
their separation that tidal effects, or effects due to their finite
size, can be ignored.  For inspiralling compact binaries, this is
believed to be a good approximation until the final few orbits.  This
amounts to replacing a perfect fluid stress-energy tensor,
\begin{equation}
T^{\mu\nu}   =(\rho +p) u^\mu u^\nu +pg^{\mu\nu} \,,
\end{equation}
with that of  ``point-masses'':
\begin{equation}
T^{\mu\nu}   = \sum_A m_A \delta^3 ({\bf x} - {\bf x}_A ) u^\mu u^\nu
/u^0 \sqrt{-g} \,.
\end{equation}
However, because of gravitational 
non-linearities, such a stress-energy tensor will
lead to infinities at the location of each body, 
hence one must find a way to regularize in order to
isolate the physically relevant terms.
Blanchet {\it et al.} use a regularization procedure based on the
Hadamard ``partie fini''.  Our approach, which is less formal, though
probably equivalent, is to isolate those terms in any integral of
fields over a body that neither vanish nor blow up as $D$, the size
of the body, shrinks to zero. 
Terms that vanish as $D^N$ represent tidal and spin effects (and their
relativistic generalizations), which we are ignoring.  Terms that
diverge as $D^{-N}$ are ``self-energy'' terms; we assume that these
can be uniformly absorbed into renormalized masses for the bodies.  It
is important to stress that this is an assumption, whose validity has
been checked in general only to 1PN order (no Nordtvedt effect in GR)
and under restricted circumstances to 2PN order.  The result is a
well-defined procedure for keeping ``finite, point-mass'' terms.

With these assumptions, the equations of motion for each body take the
form of a geodesic equation, 
\begin{equation}
\rho^* {d^2x^j \over dt^2} + \rho^* \Gamma_{\mu\nu}^j v^\mu v^\nu -
\rho^* \Gamma_{\mu\nu}^0 v^j v^\mu v^\nu =0 \,,
\label{geodesic}
\end{equation}
where $\rho^* = \rho u^0 \sqrt{-g} \to \sum_A m_A \delta^3 ({\bf x} -
{\bf x}_A )$.  

To obtain equations of motion valid through 3.5PN order, it is
necessary to iterate the relaxed Einstein equation four times.
Evaluating the resulting Poisson-like potentials for two fluid balls,
integrating the equation of motion (\ref{geodesic}) over one of the bodies, and
keeping only terms that are finite as the bodies shrink in size, one
obtains equations of motion of the schematic form
\begin{subeqnarray}
a_1^i &=& -{{m_2 } \over r^2}  [ n^i + O(\epsilon)
+ O(\epsilon^{2}) + O(\epsilon^{5/2}) + O(\epsilon^{3}) +
O(\epsilon^{7/2}) + \dots ] \,, \\
a_2^i &=& {{m_1 } \over r^2} [ 1 \rightleftharpoons 2 ] \,, 
\end{subeqnarray} 
where $r$ is
the distance
between the bodies and $n^i \equiv (x_1^i-x_2^i)/r$.
The expansion parameter $\epsilon$ is related to the orbital variables
by $\epsilon \sim m/r \sim v^2$, $v$ is the relative velocity, 
and $m=m_1 + m_2$ is
the total mass ($G=c=1$).  

We have evaluated all contributions to $h^{\mu\nu}$ formally through
3.5PN order in terms of Poisson-like potentials, 
and have calculated them explicitly for two compact
bodies through 2.5PN order and at 3.5PN order.  The formidable task of
evaluating the contributions at 3PN order is in progress.  At 2PN
order, we obtain equations of motion in complete agreement with those
of Damour and Deruelle (Eqs. (154) - (160) of Ref. \citen{damour300}) 
and Blanchet {\it et al.}
(Eq. (8.4) of Ref. \citen{lucfayeponsot}).  

The contributions at 2.5PN and 3.5PN order represent
gravitional-radiation reaction and its post-Newtonian corrections.
Iyer and Will \cite{iyerwill} 
have shown that, assuming energy and angular momentum
balance, the relative two-body equations of motion at 2.5PN order can
be written in the form 
\begin{equation}
{\bf a}=-{8 \over 5} \eta (m/r^2)(m/r)\bigl[-(A_{2.5}+A_{3.5})\dot r
{\bf n} + (B_{2.5}+B_{3.5}){\bf v}\bigr] \,,
\end{equation}
where $\eta=m_1m_2/m^2$, $\dot r = dr/dt$, and 
\begin{subeqnarray}
A_{2.5} &=& (3+3\beta)v^2+({23 \over 3} +2\alpha-3\beta){m \over r} -
5\beta {\dot r}^2 \,,\\
B_{2.5} &=& (2+\alpha)v^2 + (2-\alpha){m \over r} -3(1+\alpha){\dot
r}^2 \,, 
\end{subeqnarray}
where $\alpha$ and $\beta$ are arbitrary, and reflect the effects of
coordinate freedom on the equations of motion.  The values
$\alpha=4$, $\beta=5$ correspond to the so-called ``Burke-Thorne''
gauge, in which the radiation reaction is expressed solely as a
quasi-Newtonian potential 
$\Phi_{RR} = - \frac{1}{5} M_{ij}^{(5)}(t) x^ix^j $, where
$M_{ij}(t)$ is the traceless moment of
inertia tensor of the system and the superscript $(5)$ denotes five
time derivatives.  This also corresponds to the gauge used
by Blanchet \cite{lucreaction}.  Our 2.5PN equations of motion yield the values
$\alpha=-1$, $\beta=0$, which corresponds to the gauge used by Damour
and Deruelle (Eq. (161) of Ref. \citen{damour300}).  

At 3.5PN order,
the expressions for $A_{3.5}$ and $B_{3.5}$ are 
\begin{subeqnarray}
A_{3.5} &=&a_1v^4+a_2v^2m/r+a_3v^2\dot r^2+a_4\dot r^2m/r
+a_5\dot r^4+a_6(m/r)^2 \,, \\
B_{3.5} &=&b_1v^4+b_2v^2m/r+b_3v^2\dot r^2+b_4\dot r^2m/r
+b_5\dot r^4+b_6(m/r)^2 \,.
\end{subeqnarray} 
Energy and angular momentum balance yield values for 
the 12 coefficients modulo 6 arbitrary gauge parameters.  Our 3.5PN
equations of motion yield the values
\begin{subeqnarray}
a_1 &=&-{183 \over 28} - {15 \over 2} \eta \,,\quad b_1 = -{313 \over
28}-{3 \over 2} \eta \,, \\
a_2 &=&-{173 \over 14} - {186 \over 6} \eta \,,\quad  b_2 = {205 \over
42}+{37 \over 2} \eta \,, \\
a_3 &=&{285 \over 4} + {15 \over 2} \eta \,, \quad \quad b_3 = {339 \over
4}+{3 \over 2} \eta \,, \\
a_4 &=&-{147 \over 4} - {47} \eta \,, \quad b_4 = -{205 \over
12}-{106 \over 3} \eta \,, \\
a_5 &=& -70\,, \quad \quad \quad \quad b_5 = -75 \,, \\
a_6 &=&-{989 \over 14} - {23} \eta \,, \quad b_6 = -{1325 \over
42}-{13} \eta \,. 
\end{subeqnarray}
By comparing these values with the Iyer and Will expressions
(Eqs. (2.18) of Ref. \citen{iyerwill}), we 
obtain values for the 6 arbitrary gauge parameters.  The fact that 12
constraints yield a consistent solution for the 6 parameters is a
useful check of the method and algebra.  The result yields radiation
reaction equations to 3.5PN order in the generalization 
of Damour-Deruelle gauge.
Blanchet's multipole expressions \cite{lucreaction} 
for radiation reaction at 3.5PN
order yield a different set of coefficients, which correspond to
the generalization of Burke-Thorne gauge.

\section{Gravitational Radiation Waveform and Energy Flux}

By iterating the relaxed Einstein equations for field points in the
far zone, and substituting the two-body equations of motion to the
appropriate order, we obtain 
an explicit formula for 
the 
transverse-traceless (TT) part of the radiation-zone field, denoted
$h^{ij}$, which is the waveform to be detected in laser
interferometric systems.  In terms of an expansion
beyond the quadrupole formula, it has the schematic form,
\begin{equation}
h^{ij} = {{2\mu} \over R} 
\left\{ \tilde Q^{ij} [ 1 + O(\epsilon^{1/2}) + O(\epsilon)
+ O(\epsilon^{3/2}) + O(\epsilon^2) \dots ] \right\} _{TT} \,,
\label{1-1}
\end{equation}
%label{1-1}
where $\mu$ is the reduced mass, $R$ is the distance to the source, 
and  $\tilde Q^{ij}$ represents two time
derivatives of
the mass quadrupole moment tensor (the series actually contains
multipole orders beyond quadrupole).
The 0.5PN and 1PN terms were derived
by Wagoner and Will,\cite{wagwill} the 1.5PN terms by Wiseman.\cite{magnum}  
The contribution of
gravitational-wave ``tails'', caused by backscatter of the outgoing
radiation off the background spacetime curvature, at
$O(\epsilon^{3/2})$, were derived and 
studied by several authors.
The 2PN terms including 2PN tail contributions 
were derived by two independent groups and 
are in complete agreement (see Refs. \citen{opus} and \citen{bdi2pn} for details
and references to earlier work).  The 2.5PN terms and various specific
higher-order terms, such as ``tails-of-tails'' have been derived by
Blanchet and collaborators, while the formidable job of evaluating the
3PN terms is still in progress.

To illustrate some of the results of the DIRE method, we note that, in
calculating the gravitational waveform 
through 2PN order, we explicitly retained all terms that depend on positive
powers of the
radius $\cal R$ of the near zone.\cite{opus} \ The multipole
expansion of the integral over the
near zone yielded
\begin{eqnarray}
 h_{\cal N}^{ij}(t,{\bf x})
&=& {\cal R}{\rm -\, independent \; terms} 
+ 1 /{\cal R} \; {\rm terms} \nonumber \\
&&- {1912 \over 315} {m \over R}\; ^{(4)}{Q^{ij}}(u) {\cal R} \,,
\end{eqnarray}
where $^{(4)}{Q^{ij}}(u)$ represents four time derivative of the
quadrupole moment of the source.  The $\cal R$-dependent term is of
2PN order; this explains why sweeping ``divergent'' terms under the rug at 1PN
order was successful, if not fully justified.
The integral over the rest of the null cone yields
\begin{eqnarray}
h_{{\cal C}-{\cal N}}^{ij}(t,{\bf x}) &=&
{{4m} \over R} \int_0^\infty ds \;
^{(4)}{Q^{ij}}(u-s) \left [ \ln \left ( {s \over
{2R+s}} \right ) + {11 \over 12} \right ] \nonumber \\
&&+  {{4m} \over 3R} \hat N^k \int_0^\infty ds \;
^{(5)}{Q^{ijk}}(u-s) \left [ \ln \left ( {s \over
{2R+s}} \right ) + {97 \over 60} \right ] \nonumber \\
&&-  {{16m} \over 3R} \epsilon^{(i|ka} \hat N^k \int_0^\infty ds \;
^{(4)}{J^{a|j)}}(u-s) \left [ \ln \left ( {s \over
{2R+s}} \right ) + {7 \over 6} \right ] \nonumber \\
&&+ {1912 \over 315} {m \over R}\; ^{(4)}{Q^{ij}}(u) {\cal R} \,.
\end{eqnarray}
Notice that the $\cal R$-dependent term from the outer integral
exactly cancels that from the near-zone integral.
The other terms, involving integrals over the past history of the
source are the ``tails'', and are in complete agreement with tail
terms derived by other methods.  In addition by suitably combining the 
tail terms with the leading order quadrupole ($Q^{ij}$), octopole
($Q^{ijk}$), and current quadrupole ($J^{aj}$) terms in the waveform,
one can show that part of the effect of these terms, say for a
circular orbit of frequency $\omega$,  is to convert the
phase of the wave from $\psi = \omega (t-R)$ to $\psi = \omega (t-R
-2m \ln R - {\rm const})$.  This demonstrates that, despite the use of
a flat-spacetime wave equation to embody Einstein's equations, the
propagation of the radiation follows the true, Schwarzschild-like null
cones far from the source.

There are also contributions to the waveform
due to intrinsic spin of the bodies, which, for compact bodies, occur 
at $O(\epsilon^{3/2})$ (spin-orbit) and
$O(\epsilon^2)$ (spin-spin); these have been 
calculated elsewhere.\cite{kww}

Given the gravitational waveform, one can 
compute the rate at which energy is carried off by the radiation
(schematically $\int \dot h \dot h d\Omega$,
the gravitational analog of the Poynting
flux).  
{}For the special case of 
non-spinning bodies moving on quasi-circular orbits ({\it i.e.}
circular apart from a slow inspiral), the energy flux through 2PN
order has the 
form
\begin{eqnarray}
{dE \over dt} &= &{32 \over 5} \eta^2 {\left ({m \over r} \right )}^5
\biggl [ 1 - {m \over r} \left ( {2927 \over 336} + {5 \over 4} \eta
\right )
\nonumber \\
&&+ 4\pi{\left ( {m \over r} \right )}^{3/2} 
+ {\left ( {m \over r} \right )}^2 \left (
{293383 \over 9072} + {380 \over 9} \eta \right ) \biggr ] \;,
\label{edot}
\end{eqnarray}
%label{edot}
The first term is the quadrupole
contribution, the second term is the 1PN contribution, the
third term, with the coefficient $4\pi$, 
is the ``tail'' contribution,
and the fourth term is the 2PN
contribution first reported jointly by Blanchet {\it et al.}\cite{bdiww}

Similar expressions can be derived for the loss of angular momentum
and linear momentum.  These losses react
back on the orbit to circularize it and cause it to inspiral.  Radiation
of linear momentum can also cause ``radiation recoil'', a phenomenon
which is probably unobservable.  The
result is that the orbital phase (and consequently the 
gravitational-wave 
phase) evolves non-linearly with time.  It is the sensitivity of the
broad-band LIGO and VIRGO-type detectors to phase that makes the  
higher-order contributions to $dE/dt$ so observationally relevant.
A ready-to-use set
of formulae for the 2PN gravitational waveform template, including the
non-linear evolution of the gravitational-wave frequency (not
including spin effects) have been published\cite{biww}  and
incorporated into the Gravitational Radiation Analysis and Simulation
Package (GRASP), a publically available software toolkit.\cite{grasp}

\section*{Acknowledgments}

This work was supported in part by the National Science Foundation,
Grant Number PHY 96-00049.


\begin{thebibliography}{}

\bibitem{snowmass} K. S. Thorne, in {\it Proceedings of the Snowmass
95 Summer Study on Particle and Nuclear Astrophysics and Cosmology},
edited by E. W. Kolb and R. Peccei (World Scientific, Singapore, 1995), 
p. 398 (gr-qc/9506086).
\bibitem{finnchern} L. S. Finn and D. F. Chernoff, Phys. Rev. {\bf
D47} (1993), 2198 (gr-qc/9301003). \\
C. Cutler and \'E. E. Flanagan, Phys. Rev. 
{\bf D49} (1994), 2658 (gr-qc/9402014). \\
E. Poisson and C. M. Will, Phys. Rev. {\bf D52} (1995), 848 (gr-qc/9502040).
\\
T. Damour, B. R. Iyer and B. S. Sathyaprakash, Phys. Rev. {\bf D57}
(1998), 885. \\
E. Poisson, in {\it Proceedings of the Second
Gravitational-Wave Data Analysis Workshop}, in press.
\bibitem{wagwill} R. V. Wagoner and C. M. Will, Astrophys. J. {\bf 210} (1976),
764.
\bibitem{convention}
This convention holds sway, 
despite the fact that pure Newtonian gravity predicts no
gravitational radiation. It is often the source of confusion, since
the energy carried by the lowest-order ``Newtonian'' quadrupole
radiation manifests itself in a (post)$^{5/2}$-Newtonian, or $O[(v/c)^5]$
correction in the equation of motion.
\bibitem{walkerwill} M. Walker and C. M. Will, Phys. Rev. Lett. {\bf
45} (1980), 1741. \\
J. L. Anderson, 
Phys. Rev. Lett. {\bf 45} (1980), 1745. \\
T. Damour, Phys. Rev. Lett. {\bf 51} (1979), 1019
(1983). \\
D. Christodoulou and B. G. Schmidt, Commun. Math. Phys.
{\bf 68} (1979), 275. \\
R. A. Isaacson, J. S. Welling, and J. Winicour, Phys.
Rev. Lett. {\bf 53} (1984), 1870.
\bibitem{bd86}  L. Blanchet, T. Damour, Phil. Trans. R. Soc. London
{\bf A320} (1986), 379. \\
L. Blanchet and T. Damour, Phys. Rev. D {\bf 37} (1988), 1410. \\
L. Blanchet and T. Damour, Ann. Inst. H. Poincar\'e
(Phys. Theorique) {\bf 50} (1989), 377. \\
T. Damour and B. R. Iyer,  Ann. Inst. H. Poincar\'e
(Phys. Theorique) {\bf 54} (1991), 115. \\
L. Blanchet, Phys. Rev. {\bf D51} (1995), 2559 (gr-qc/9501030). \\ 
L. Blanchet and T. Damour, Phys. Rev. {\bf D46}
(1992), 4304.
\bibitem{thorne80} K. S. Thorne, Rev. Mod. Phys. {\bf 52}
(1980), 299.
\bibitem{ew} R. Epstein and R. V. Wagoner, Astrophys. J. {\bf 197} (1975),
717 (EW).
\bibitem{tts96}  E. Poisson, Phys. Rev.  {\bf D47} (1993), 1497. \\
E. Poisson and M. Sasaki, Phys. Rev. {\bf D51}
(1995), 5753 (gr-qc/9412027). \\
C. Cutler, L. S. Finn, E. Poisson, and G. J. Sussman,
Phys. Rev. {\bf D47} (1993), 1511. \\
M. Sasaki, Prog. Theor. Phys. {\bf 92} (1994), 17 (gr-qc/9402042). \\
H. Tagoshi and M. Sasaki, Prog. Theor. Phys. {\bf
92} (1994), 745 (gr-qc/9405062). \\
T. Tanaka, H. Tagoshi, and M. Sasaki, Prog. Theor.
Phys. {\bf 96} (1996), 1087 (gr-qc/9701050).
\bibitem{magnum} A. G. Wiseman, Phys. Rev. {\bf D46} (1992), 1517.
\bibitem{opus} C. M. Will and A. G. Wiseman, Phys. Rev. {\bf D54} (1996),
4813 (gr-qc/9608012).
\bibitem{christo} A. G. Wiseman and C. M. Will, Phys. Rev. {\bf D44} (1991),
R2945.
\bibitem{willpati}  M. E. Pati and C. M. Will, papers in
preparation.
\bibitem{damour300} T. Damour , 
1987, in {\it 300 Years of Gravitation}, eds Hawking S W and 
Israel W, 128--198 (Cambridge University Press, Cambridge). 
\bibitem{lucfayeponsot} L. Blanchet, G. Faye and B. Ponsot, Phys.
Rev. {\bf D58} (1998), 124002  (gr-qc/9804079).
\bibitem{iyerwill} B. Iyer and C. M. Will, 
Phys. Rev. {\bf D52} (1995), 6882.
\bibitem{lucreaction} L. Blanchet, Phys. Rev. {\bf D47} (1993),
4392.
\bibitem{bdi2pn}  L. Blanchet, T. Damour, and B. R. Iyer, Phys. Rev. 
{\bf D51} (1995), 5360 (gr-qc/9501029).
\bibitem{kww} L. E. Kidder, C. M. Will, and A. G. Wiseman, Phys. Rev.
{\bf D47} (1993), R4183 (gr-qc/9211025);
L. E. Kidder, Phys. Rev.  {\bf D52} (1995), 821 (gr-qc/9506022).
\bibitem{bdiww} L. Blanchet, T. Damour, B. R. Iyer, C. M. Will and A.
G. Wiseman, Phys. Rev. Lett. {\bf 74} (1995), 3515 (gr-qc/9501027).
\bibitem{biww} L. Blanchet, B. R. Iyer, C. M. Will, and A. G. Wiseman,
Class. Quantum Grav. {\bf 13} (1996), 575 (gr-qc/9602024). 
\bibitem{grasp} See
http://www.ligo.caltech.edu/LIGO\_web/Collaboration/manual.pdf.

\end{thebibliography}
\end{document}